
\magnification=1200
\font\titlea=cmb10 scaled\magstep1

\baselineskip=12pt
\tolerance = 10000
\rightline{IC-92-226}
\rightline{hepth@xxx/9208065}
\baselineskip=18pt
\vskip .5cm
\centerline{\titlea The Lax Operator Approach for the Virasoro and}
\centerline{\titlea the W-Constraints in the Generalized KdV Hierarchy}
\vskip 2cm
\baselineskip=14pt
\centerline{Sudhakar Panda$\,^\ast$}
\bigskip
\centerline{and}
\bigskip
\centerline{Shibaji Roy}
\bigskip
\centerline{\it International Centre for Theoretical Physics}
\centerline{\it Trieste - Italy}
\vskip 1.0cm
\baselineskip=18pt
\centerline{\titlea Abstract}
\bigskip
We show directly in the Lax operator approach how the Virasoro and
W-constraints
on the $\tau$-function arise in the $p$-reduced KP hierarchy or Generalized
KdV hierarchy. In particular, we consider the KdV and the Boussinesq hierarchy
to show that the Virasoro and the W-constraints follow from the string equation
by expanding the \lq\lq additional symmetry" operator in terms of the Lax
operator. We also mention how this method could be generalized for higher
KdV hierarchies.
\vfill
\hbox to 3.5cm {\hrulefill}\par
\item{($\ast$)} Address after 1st November 1992: Institute for Theoretical
Physics, Groningen University, 9747 AG Groningen, The Netherlands.
\eject
{\titlea I. Introduction:}
\bigskip
It is by now a well recognized fact that the integrable models play very
interesting
role in the matrix model formulation of two dimensional (2D) quantum gravity
[1-5],
2D topological gravity [6-8] and the intersection theory on the moduli space of
Riemann surfaces [9-11]. In the matrix model approach of the 2D gravity, one
employs the method of orthogonal polynomials and makes use of the operators
$Q$ and $P$ corresponding to the insertions of the spectral parameter and
a derivative with respect to it in the matrix integral [12,13]. As Douglas
argued, $Q$ and $P$ can be realized in terms of some finite order differential
operators and can be recognized as the Lax-pair of an associated integrable
hierarchy [14-16]. Since the operators $P$ and $Q$ are conjugate to each other,
they satisfy the so-called \lq\lq string equation" [13] $[P, Q]~=~1$. Once the
pair of operators ($P$,$Q$) is recognized as the Lax pair and we set their
commutator to be one, it puts a very stringent condition on the coefficient
functions of the Lax operator. This in turn implies an infinite number of
additional symmetries of the integrable hierarchies and can be recognized as
the Virasoro and W-constraints. In the usual integrable models, the symmetries
arise from the isospectral deformation of the Lax operator, but these
additional \lq\lq time"-dependent symmetries originate from a general
Galilean transformation of the evolution parameters as emphasized in [17]. The
origin and the geometry of the string equation and its connection with the
Sato-Grassmanian can be found in the recent literature [18-21].

By formulating the Hermitian one matrix model [22] and two matrix
model [23], with specific interaction, in terms of the continuum
Schwinger-Dyson
equations it is shown that they give rise to a semi-infinite set of Virasoro
(for 1-matrix model) and $W^{(3)}$-constraints (for 2-matrix model) on the
square root of the partition function. Through an identification of the square
root of the partition function with the $\tau$-function of the corresponding
integrable hierarchy it has been conjectured [22] that the whole set of
Virasoro and W-constraints follow as a consequence of the string equation
itself.
By making use of the string equation and the associated biHamiltonian structure
of the KdV hierarchy it has been shown (although in a very indirect way) that
this is indeed true [7]. Goeree [24] has also shown using the vertex operator
techniques of KP hierarchy developed in [25,26], that not only the Virasoro
constraints but also the W-constraints follow from the string equation.
However, the additional symmetry operator $M$ as used there, does not
reproduce the correct W-constraints and needed to be modified.
This has been reported by us in a recent letter [27].

Taking into account the above modification and to make the structure more
transparent we develop a direct approach in this paper to show how the
generators of the additional symmetries [28-31] give rise to the Virasoro and
W-constraints. We use the method of ($L, M$) pair of the $p$-reduced KP
hierarchy
by which one can construct the generators of the additional symmetries
associated
with such integrable systems. We then expand the operator $M$ as a power series
of
the Lax operator. In this way the residues of the generators of the additional
symmetries can be
recognized as the Virasoro and the W-constraints found in the matrix model
approach.

The paper is organized as follows. In section II, we consider the $2$-reduced
(KdV) KP hierarchy and explain our method. The $3$-reduced (Boussinesq) KP
hierarchy
is considered in details in section III. In section IV, we discuss the
generalization
of the method for higher KdV hierarchy. Our conclusions are drawn in Section V.
\vskip .5cm
{\titlea II. KdV Hierarchy and the Virasoro Constraints:}
\bigskip
The $2$-reduced KP hierarchy or KdV hierarchy is described in terms of the
following Lax equation,
$$
{\partial L \over {\partial t_{2k+1}}}~=~[ L^{{2k+1\over 2}}_+~,~L ]
{}~~~~~~~~~~k~=~0,1, 2, 3,\ldots
\eqno(2.1)
$$
Here, $L~=~{\partial^2\over \partial x^2}~ +~ u (x,t_3,t_5,\ldots)~\equiv~
\partial^2~+~u (x,t)$ is the Lax operator of the KdV hierarchy and
$L^{{2k+1\over 2}}_+$
is the non-negative differential part of the $(2k+1)$th power of the formal
pseudo-differential operator $L^{{1\over 2}}$. Together they are known as the
Lax-pair. $L^{{1\over 2}}$ is the two reduced KP Lax operator and has a formal
expansion in the form
$$
\eqalign{L^{{1\over 2}}~=&~\partial + {u\over 2} \partial^{-1} -
{u^{\prime}\over 4} \partial^{-2} + ({u^{\prime\prime}\over 8} -
{u^2\over 8}) \partial^{-3} + ({3 u u^{\prime}\over 8} -
{u^{\prime\prime\prime}
\over 16})\partial^{-4}\cr &\qquad \qquad+ ({u^{\prime\prime\prime\prime}
\over 32}
- {11 {u^\prime}^2\over 32} + {u^3\over 16} - {7\over 16} u u^{\prime\prime})
\partial^{-5} + \cdots\cr}
\eqno(2.2)
$$
such that we have
$$
( L^{1\over 2} )~=~L~=~\partial^2~+~u
\eqno(2.3)
$$
Here, \lq prime\rq  denotes the differentiation with respect to $x$. $t_{2k+1},
{}~k=0,1,2,\ldots$ are the infinite number of evolution parameters associated
with
the KdV hierarchy. From (2.1), one can identify $t_1 \equiv x$. Also note that
the differential part of the even powers of $L^{1\over 2}$ will commute with
$L$. In order to evaluate various powers of $L^{1\over 2}$, one makes use of
the Liebnitz rule
$$
\partial^{-i} f~=~\sum_{j=0}^\infty (-)^j {i+j-1\choose j} f^{(j)}
\partial^{-i-j}
\eqno(2.4)
$$
where we have denoted $f^{(j)}~=~{\partial^j f\over \partial x^j}$.

According to Douglas, the string equation [13] (for one matrix model)
corresponding
to $k$-th critical point is given by
$$
[ L^{{2k+1\over 2}}_+, L ]~=~1
\eqno(2.5)
$$
An arbitrary massive model which interpolates between various critical points
can be written by generalizing (2.5) as follows,
$$
\sum_{k=1}^\infty ( k + {1\over 2}) t_{2k+1}~ [ L, L^{{2(k-1)+1}
\over 2}_+]~ = ~1
\eqno(2.6a)
$$
where we have introduced an infinite number of evolution parameters $t_{2k+1}$
proportional to $-1/ (k+{1\over 2})$. Note that (2.6a) can also be
expressed in an equivalent form as
$$
\sum_{k=1}^\infty (k+{1\over 2}) t_{2k+1}~res~L^{{2(k-1)+1\over 2}}
+ {1\over 2} x~=~0
\eqno(2.6b)
$$
We have integrated (2.6a) once with respect to $x$ in order to derive (2.6b).
Also ``$res$" here simply means the coefficient of $\partial^{-1}$ term in
the pseudo-differential operator. Equation (2.6) can be written
in a different form given by
$$
[~L, ( M L^{-{1\over 2}})_+~]~=~1
\eqno(2.7)
$$
The operator $M$ for the 2-reduced KP hierarchy is defined as
$$
M~\equiv ~{1\over 2}~K~( \sum_{n=1\atop n\neq 0 (mod 2)}^{\infty} n t_n
\partial^{n-1} )~K^{-1}
\eqno(2.8)
$$
We would like to point out here that it not necessary to remove the coordinates
$t_n$ where $n = 0(mod 2)$ in the definition of $M$ in order to get the
correct Virasoro constraints. But, this becomes necessary in order to get right
$W$-constraints [27]. We,
therefore, define $M$ in this way from the beginning so that we do not face
any problem later. Also, here,
 $K~=~1 + \sum_{i=1}^{\infty} a_i (x,t)~\partial^{-i}$ is a
pseudo-differential operator known as the Zakharov-Shabat dressing operator
and satisfies the relation [25]
$$
L^{{1\over 2}}~ =~ K \partial K^{-1}
\eqno(2.9)
$$
This fixes the coefficients of $K$ in terms of $u(x,t)$ and their derivatives.
Using (2.9), we rewrite $M$ as
$$
M~=~{1\over 2}~K x K^{-1} + {1\over 2} \sum_{n=3\atop n\neq 0 (mod 2)}^\infty n
t_n L^{{n-1\over 2}}
\eqno(2.10)
$$
Thus, we have
$$
(M L^{-{1\over 2}})_+~=~{1\over 2} \sum_{n=3\atop n\neq 0 (mod 2)}^\infty n t_n
L^{{n-2\over 2}}_+
\eqno(2.11)
$$
Substituting (2.11) into (2.7) we recover (2.6a). So, (2.7) is indeed an
equivalent
form of (2.6) i.e. Douglas' string equation. Using the definition of $M$ in
(2.10) we can show that [30]
$$
[~L^{{1\over 2}}, M~]~=~{1\over 2}
\eqno(2.12)
$$
and therefore,
$$
[~L, M L^{-{1\over 2}}~]~=~1
\eqno(2.13)
$$
In view of the string equation (2.7), one concludes that $(ML^{-{1\over 2}})_-$
which is purely pseudo-differential part of $ML^{-{1\over 2}}$ should commute
with $L$. Using (2.12) one can derive that
$$
[~M, L^{-{1\over 2}}~]~=~{1\over 2} L^{-1}
\eqno(2.14)
$$
Since $(ML^{-{1\over 2}})_-$ commutes with $L$ and it satisfies (2.14),
therefore, it must be proportional to $L^{-1}$. We set,
$$
(M L^{-{1\over 2}})_-~=~\alpha L^{-1}
\eqno(2.15)
$$
where $\alpha$ is an arbitrary constant which can not be determined just from
the string equation (2.7) as mentioned in ref.[24].
{}From this it follows that for $n\geq 0$ we have
$$
(M L^{n+{1\over 2}})_-~=~(( ML^{-{1\over 2}})_-~L^{n+1} )_-~=~0
\eqno(2.16)
$$
The second expression is because of the fact that $L^n$ does not contain any
negative power of $\partial$ for $n\geq 0$. It has been noted in ref.[31]
that the particular combination of $L$ and $M$ (2.15) and (2.16) are the
generators of the additional symmetries of the KdV hierarchy in the sense that
they satisfy, for $n\geq -1$,
$$
{\partial L \over {\partial t_{2n+1, 1}}}~=~ [ L, ( M L^{n+{1\over 2}} )_- ]~
=~0
\eqno(2.17)
$$
These flows commute with the original KdV hierarchy flows given in (2.1),
but they do not commute among themselves and have nice interpretation in terms
of the Sato-Grassmannian [31].

We first show that (2.15) does indeed imply the string equation (2.6b) and
then work out the consequences of (2.16). The operator $M$ has an expansion
in the power series of the Lax operator in the form [32] (see appendix)
$$
M~=~{1\over 2} \sum_{n=1\atop n\neq 0 (mod2)}^\infty n t_n L^{{n-1\over
2}}~+~{1\over 2}
\sum_{i=1}^\infty V_{i+1} (x,t) L^{{-i-1\over 2}}
\eqno(2.18)
$$
The functions $V_{i+1} (x,t)$ can be be determined in terms of the coefficient
functions of the dressing operators $K$ and $K^{-1}$ as follows
$$
V_{i+1} (x,t)~=~-(i a_i~+~\sum_{j=1}^{i-1} j a_j {\tilde a}_{i-j})
\eqno(2.19)
$$
The operator $K^{-1}$ is chosen to have the form
$$
K^{-1}~=~1 + \sum_{i=1}^\infty {\tilde a}_i (x,t) \partial^{-i}
\eqno(2.20)
$$
By requiring $K K^{-1}~= 1$, ${\tilde a}_i$'s can be fixed in terms of
$a_i$'s from the relation,
$$
a_i + {\tilde a}_i + \sum_{k=1}^i \sum_{j=1}^{k-1} (-)^{i-k} {i+j-k-1\choose
i-k} a_j {\tilde a}_{k-j}^{(i-k)}~=0~~~~~~i=1,2,3,\ldots
\eqno(2.21)
$$
Taking the residue of (2.15) we get
$$
res~(M L^{-{1\over 2}})~=~0
\eqno(2.22)
$$
Inserting the expression of $M$ as given in (2.10) in the above we obtain
$$
{1\over 2} x~+ {1\over 2} \sum_{n\geq 3\atop n\neq 0 (mod 2)} n t_n res
L^{{n-2\over 2}}~=~0
\eqno(2.23)
$$
This is precisely equation (2.6b). This, therefore, establishes the
equivalence between (2.6),
(2.7) and (2.15). One defines the $\tau$-function of the KdV hierarchy as
a function depending on the coefficient functions of the Lax operator
(in the present case $u$) and their derivatives and is given by (see appendix
for the derivation),
$$
res~L^{{2k+1\over 2}}~=~{\partial\over \partial x}~{\partial\log\tau\over
\partial t_{2k+1}}
\eqno(2.24)
$$
Now, using (2.24) in (2.23) and performing an integration with respect to
$x$, and multiplying by $\tau$ we find that
$$
L_{-1}~\tau~=~0
\eqno(2.25)
$$
where we have defined the operator $L_{-1}$ as
$$
L_{-1}~\equiv~\sum_{k=1}^\infty (k+{1\over 2}) t_{2k+1} {\partial\over
\partial t_{2(k-1)+1}}~+ {1\over 4} x^2
\eqno(2.26)
$$
Thus, (2.25) can also be called as the string equation since it is equivalent
to (2.6). Next, we can work out the residue of $ML^{n+{1\over 2}}$ for
$n\geq 0$ by using the expression for $M$ as in (2.18). Note that for these
cases, the last term in (2.18) will contribute for odd values of $i$. Thus,
we need to express the functions $V_{i+1}$ in terms of the $\tau$-function.
In (2.19), we have expressed $V_{i+1}$ in terms of $a_i$ and ${\tilde a}_i$
which are some meromorphic functions and can be expressed in terms of
$\tau$-function. It can be shown that $V_{i+1}$ has the following form (see
appendix)
$$
V_{i+1}~=~-i\sum_{\alpha_1+3\alpha_3+5\alpha_5+.....=i} (-)^{\alpha_1+\alpha_3
+\alpha_5+.....} {(\partial_{t_1})^{\alpha_1}\over \alpha_1!} {({1\over 3}
\partial_{t_3})^{\alpha_3}\over \alpha_3!} {({1\over
5}\partial_{t_5})^{\alpha_5}
\over \alpha_5!}.....\log\tau
\eqno(2.27)
$$
After integrating $res~(M L^{{1\over 2}})~=0$ once with respect to $x$ and
multiplying by $\tau$, we obtain for $n=0$ that
$$
\left[\sum_{k=0}^\infty (k+{1\over 2})t_{2k+1} {\partial\over \partial
t_{2k+1}}~+~C \right]~\tau~=~0
\eqno(2.28)
$$
where $C$ is an arbitrary integration constant which appears here unlike in
(2.26) is because the scaling dimension of (2.28) is zero. This constant will
be fixed later. For $n\geq 1$, the residue of
$M L^{n+{1\over 2}}$ could also be calculated in an identical way and the
result is that
$$
\left[\sum_{k=0}^\infty (k+{1\over 2}) t_{2k+1}
{\partial\over \partial t_{2(n+k)+
1}} + {1\over 4}\sum_{i+j=2n\atop i,j\neq 0(mod2)} {\partial\over \partial t_i}
{\partial\over \partial t_j}~\right]~\tau~=~0
\eqno(2.29)
$$
The appearance of the second term in (2.29) comes from the contribution of
$V_{i+1}$ functions contained in $M$. In fact, it can be shown for $n\geq 0$
that these functions satisfy the following relation;
$$
\sum_{k=0}^n V_{2k+2} ~res~ L^{{2n-2k-1\over 2}} = {\partial \log\tau\over
\partial t_{2n+1}} + {1\over 2}{\partial\over \partial x}~\left[{1\over \tau}
\sum_{i+j=2n\atop i,j\neq 0(mod 2)} {\partial\over \partial t_i}{\partial\tau
\over \partial t_j}~\right]
\eqno(2.30)
$$
where we have used the simple identity
$$
{\partial^2\log\tau\over \partial t_i \partial t_j}~+~{\partial\log\tau\over
\partial t_i} {\partial\log\tau\over \partial t_j}~=~{1\over \tau}{\partial^2
\tau \over \partial t_i \partial t_j}
\eqno(2.31)
$$
To check the validity of (2.30), we give the expressions for few of $V_{2k+2}$,
for example,
$$
\eqalign{V_2&=~{\partial\log\tau\over \partial x}\cr
         V_4&=~{1\over 2} ({\partial^3\over \partial x^3} + 2{\partial\over
\partial t_3})~\log\tau\cr
V_6&=~({1\over 4!} {\partial^5\over \partial x^5} + {5\over 6} {\partial^2\over
\partial x^2} {\partial\over \partial t_3} + {\partial\over \partial t_5})
{}~\log\tau\cr
V_8&=~({1\over 6!} {\partial^7\over \partial x^7} + {\partial\over \partial
t_7} + {7\over 10} {\partial^2\over \partial x^2}{\partial\over \partial t_5}
+ {7\over 72} {\partial^4\over \partial x^4}{\partial\over \partial t_3}
+ {7\over 18}{\partial\over \partial x}{\partial^2\over \partial t_3^2})
{}~\log\tau\cr}
\eqno(2.32)
$$
and the $\tau$-function satisfies the following identities,
$$
\eqalign{&{1\over 48}{\partial^5\log\tau\over \partial x^5} - {1\over12}
{\partial^2\over \partial x^2}{\partial \log\tau\over \partial t_3}
+ {1\over 4} {\partial^3\log\tau\over \partial x^3}{\partial^2\log\tau\over
\partial x^2}~=~0\cr &{1\over 1440}{\partial^7\log\tau\over \partial x^7} -
{3\over 20} {\partial^2\over \partial x^2} {\partial \log\tau\over \partial
t_5} + {7\over 144}{\partial^4\over \partial x^4}{\partial\log\tau\over
\partial t_3} - {1\over 18}{\partial\over \partial x}{\partial^2\log\tau\over
\partial t_3^2}\cr +&{1\over 48}{\partial^5\log\tau\over \partial x^5}
{\partial^2\log\tau\over \partial x^2} + {5\over 12} {\partial^2\over
\partial x^2}{\partial\log\tau\over \partial t_3}{\partial^2\log\tau\over
\partial x^2} +{1\over 4}{\partial^3\log\tau\over \partial x^3}{\partial\over
\partial t_3}{\partial\log\tau\over \partial x} = 0\cr}
\eqno(2.33)
$$
Equations (2.32) and (2.33) would check that the relation (2.30) is true up to
$n=3$ and for higher values of $n$, it becomes more tedious.

It is customary in the matrix model to identify the operators appearing in
(2.28) and (2.29) as $L_0$ and $L_n$ respectively for the reason that they
satisfy a centerless semi-infinite Virasoro algebra. In general for arbitrary
value of $C$ in (2.28), the Virasoro algebra will not be satisfied. So if we
insist that these operators along with (2.25) satisfy the Virasoro algebra
in analogy with the matrix model result we have to fix the integration
constant $C$ to be ${1\over 16}$. In this situation, we have the standard form
of Virasoro constraints on the $\tau$-function
$$
L_n~\tau~=~0~~~~~~~~~~~~n\geq -1
$$
where
$$
\eqalign{L_{-1}~&=~\sum_{k=1}^\infty (k+{1\over 2}) t_{2k+1} {\partial\over
\partial t_{2(k-1)+1}}~+{1\over 4} x^2\cr L_0~&=~\sum_{k=0}^\infty (k+{1\over
2}) t_{2k+1} {\partial\over \partial t_{2k+1}}~+{1\over 16}\cr L_n~&=~\sum_{k=
0}^\infty (k+{1\over 2}) t_{2k+1} {\partial\over \partial t_{2(n+k)+1}} + {1
\over 4}\sum_{i+j=2n\atop i,j\neq 0(mod 2)} {\partial\over \partial t_i}
{\partial\over \partial t_j}~~~~n\geq 1\cr}
\eqno(2.34)
$$
 \vskip .5cm
 {\titlea III. Boussinesq Hierarchy and $W_3$-Constraints:}
 \bigskip
 The Boussinesq hierarchy can also be described in terms of the Lax
 equation given by
 $$
 {\partial L\over \partial t_{3n+i}} = [ L^{{3n+i}\over 3}_+, L]
 \qquad n=0,1,2,\ldots \quad and\quad i=1,2
 \eqno(3.1)
 $$
 Here the Lax operator $L$ for the three reduced KP hierarchy
 or Boussinesq hierarchy is a third order differential operator,
 $$
 L = \partial^3 + 4u\partial + (2u' + w)
 \eqno(3.2)
 $$
 where $u(x,t)$ and $w(x,t)$ are the coefficient functions
 of the Lax operator. The particular form of $L$ in (3.2) ensures that
 $L$ transforms covariantly under conformal transformation [32]. Note that
 we have an infinite number of evolution parameters $t_k$ where
 $k \neq 0(mod 3)$ because, for those values the commutator in (3.1) will
 vanish. Also, the formal pseudodifferential operator $L^{ 1\over 3}$
 has the expansion of the form
 $$
 \eqalign{ L^{1 \over 3} &= \partial + {4 \over 3} u \partial^{-1}
 + {1\over 3} (w-2 u')\partial^{-2} - {1\over 3} (w'
 -{2\over 3} u'
 + {16\over 3}u^2) \partial^{-3}\cr
 &-{1\over 3}(-{2\over 3}w'' + {8\over 3}
 u w - 16 u u')\partial^{-4}
 -{1\over 3}({1\over 3}w''' + {2\over 9}u'''' -{16\over 3} u'w\cr
 &\qquad\qquad - {16
 \over 3} u w' + {44\over 3} (u')^2 + 16 u u'' + {1\over 3} w^2 -
 {320\over 27} u^3)\partial^{-5}\cr
 &-{1\over 3}(-{1\over 9} w'''' -{2\over 9} u''''' + {70\over 9} u'' w
 +{20\over 3}u w'' + {110\over 9} u' w' - {100\over 3} u' u'' - {40\over
 3} u u'''\cr
 &\qquad\qquad - {5\over 3} w w' + {800\over 9} u^2 u' - {80\over 9} u^2 w)
 \partial^{-6} + \ldots\cr}
 \eqno(3.3)
 $$
  such that  $(L^{1\over 3})^3 = L$

The string equation for the $k$-th critical point can now be written as
$$
[ L^{{3k+i}\over 3}_+ , L ] = 1
\eqno(3.4)
$$
 In terms $u(x,t)$ and $w(x,t)$, this means
$$
\eqalign{{\partial u \over \partial t_{3k+i}} &= 0\cr
{\partial w\over \partial t_{3k+i}} &= 1\cr}~~~~~~~~~~~~~~ {k=0,1,2,\ldots
\atop i= 1,2}
\eqno(3.5)
$$
 Again $t_1$ can be identified with $x$ from (3.1) and for general
massive model we can write down the string equation in the form
$$
\sum_{k=1 \atop i=1,2}^\infty (k+{i\over 3}) t_{3k+i}~~ [ L, L^{3(k-1)+i
\over 3}_+ ] = 1
\eqno(3.6)
$$
 As in the KdV case we have introduced infinite number of
evolution parameters $t_{3k+i}$ proportional to $-1/(k+ {i\over 3})$.
In terms of the Gelfand-Dikii polynomials eq.(3.6) can be rewritten as
$$
\sum_{k=1 \atop i=1,2}^\infty 3 (k+{i\over 3}) t_{3k+i} {\partial R_k^
{(i)} \over \partial x} + 1 = 0
\eqno(3.7)
$$
 and
$$
\sum_{k=1 \atop i=1,2}^\infty 6 (k+{i\over 3})t_{3k+i} {\partial
{\tilde R}_k^{(i)}
\over \partial y} +1 = 0
\eqno(3.8)
$$
 Here we have identified $t_2$ with $y$ and also
 note that we have got two equations from the string equation
(3.6). This is because that the Gelfand-Dikii polynomials satisfy the
relation
$$
{\partial R_k^{(i)} \over \partial x} = 2 {\partial {\tilde R}_k^{(i)}
\over \partial y}~~~~~~~~~~~{k=1,2,\ldots \atop i=1,2}
\eqno(3.9)
$$
 So, (3.7) and (3.8) are really two equivalent forms of the string
equation (3.6) and we will work only with (3.7). The Gelfand-Dikii
polynomials can be calculated from the Lax equation which we write in terms
of a $2\times 2$ matrix notation as follows,
$$
{{\partial u \over \partial t_{3n+i}} \choose {\partial w \over \partial t
_{3n+i}}} = 3 K_1 { R_{n+1}^{(i)} \choose {\tilde R}_{n+1}^{(i)}}
{}~~~~~~~~~~~~ {n=0,1,2,\ldots \atop i=1,2}
\eqno(3.10)
$$
 and they satisfy the following recursion relation
$$
K_2 {R_n^{(i)} \choose {\tilde R}_n^{(i)}} = 3 K_1 {R_{n+1}^{(i)}
\choose {\tilde R}_{n+1}^{(i)}}
\eqno(3.11)
$$
 where $K_1$ and $K_2$ are the biHamiltonian structures associated
with the Boussinesq hierarchy and they are given below,
$$
K_1 = \left(\matrix{0&\qquad\partial\cr
                      \partial &\qquad 0\cr}\right)
$$
and
$$
K_2 = \left(\matrix{{1\over 2} \partial^3 + 2u\partial + u' & 3w\partial
   +2 w'\cr
   3 w\partial + w' & \eqalign{&-{2 \over 3}(\partial^5 + 20 u\partial^3 +
   30 u'\partial^2 + 18 u''\partial\cr
   &\qquad\qquad 64 u^2\partial + 4u''' + 64 u u')\cr}\cr}\right)
\eqno(3.12)
$$
 The first few Gelfand-Dikii polynomials are listed here,
$$
\eqalign{ R_0^{(1)} &= 1\cr
            R_1^{(1)} &= {1 \over 3} w\cr
            R_2^{(1)} &= -{2 \over 27} u'''' -{16\over 9} u u''
            -{8 \over 9} (u')^2 + {2\over 9} w^2 - {256\over 81} u^3\cr
           {\tilde R}_0^{(1)} &= 0\cr
           {\tilde R}_1^{(1)} &= {1\over 3} u\cr
           {\tilde R}_2^{(1)} &= {1\over 18} w'' + {4\over 9} u w\cr
           R_0^{(2)} &= 0\cr
           R_1^{(2)} &= -{2\over 9} u'' -{16\over 9} u^2\cr
           R_2^{(2)} &= -{2\over 27} w''''- {80 \over 27} u^2 w -
            {20\over 27} u w'' - {10\over 27} u' w' - {10\over 27} w u''
            \cr
          {\tilde R}_0^{(2)} &= {1\over 4}\cr
           {\tilde R}_1^{(2)} &= {1\over 6} w\cr
           {\tilde R}_2^{(2)} &= -{1\over 27} u'''' -{20\over 27} u u''
           -{5\over 9} (u')^2 +{5\over 36} w^2 -{80 \over 81} u^3\cr}
 \eqno(3.13)
$$
The Gelfand-Dikii polynomials appeared in (3.7) could be related
to the $\tau$-function of the Boussinesq hierarchy by the following relation
(see appendix for the derivation),
$$
R_k^{(i)} = (L^{{3(k-1)+i}\over 3})_{-2} + {1\over 2} (L^{{3(k-1)+i}
\over 3})'_{-1} = {1\over 2} {\partial^2 \over {\partial t_2 \partial t_
{3(k-1)+i}}} \log\tau~~~~~~~~~ k=1,2,\ldots
\eqno(3.14)
$$
 where the subscript `-2' and `-1' refer to the coefficient of
$\partial^{-2}$ and $\partial^{-1}$ terms of the expansion of the formal
pseudodifferential operators. Using (3.14) in eq.(3.7) and then integrating
with respect to $x$ and $y$ once and the multiplying by $\tau$ we get,
$$
\left[\sum_{k=1 \atop i=1,2}^\infty (k+{i\over 3}) t_{3k+i} {\partial \over
\partial t_{3(k-1)+i}} + {2\over 3} xy\right] \tau = 0
\eqno(3.15)
$$
 This, therefore, is an equivalent form of the string equation
(3.6). Following the procedure described in sec.II, it is easy to show that
eq.(3.6) can also be written in another form
$$
[ L, (M L^{-{2\over 3}})_+ ] = 1
\eqno(3.16)
$$
 where the operator $M$ is now defined as
$$
M = {1\over 3} K x K^{-1} + {1\over 3} \sum_{n=2 \atop n\not= 0(mod 3)}
^\infty n t_n L^{ n-1 \over 3}
\eqno(3.17)
$$
We note that, when $M$ is multiplied by $L^{-{2\over 3}}$, then the first
term and $n = 2$ term in the sum in (3.17) will not contribute in
$(M L^{-{2\over 3}})_+$. So, (3.16) precisely matches with (3.6). Using
the expression of $M$ in (3.17), we can show that the operators $L$ and
$M$ satisfy the following commutation relations,
$$
[ L^{1\over 3}, M ] = {1\over 3}
\eqno(3.18a)
$$
$$
[ L, M L^{-{2 \over 3}} ] = 1
\eqno(3.18b)
$$
and
$$
[ M, L^{-{2\over 3}} ] = {2\over 3} L^{-1}
\eqno(3.18c)
$$
{}From the relations (3.18) and (3.16), we conclude that
$$
( M L^{-{2\over 3}})_- = \alpha L^{-1}
\eqno(3.19)
$$
where $\alpha$ is again some arbitrary constant. The residue of eq.(3.19)
can be shown to be the string equation (3.6). As before, eq.(3.19) now
implies,
$$
( M L^{n+{1\over 3}})_- = ((M L^{-{2\over 3}})_- L^{n+1})_- = 0
\qquad \quad n\geq 0
\eqno(3.20)
$$
The expansion of $M$ in terms of Boussinesq Lax operator has the following
form
$$
M = {1\over 3}\sum_{n=1 \atop n \ne 0(mod 3)}^\infty n t_n
L^{n-1 \over 3} + {1 \over 3} \sum_{ i=1}^\infty V_{i+1} L^{-i-3 \over 3}
\eqno(3.21)
$$
The functions $V_{i+1}$ can be expressed in terms of the $\tau$ function
of Boussinesq hierarchy and is given as,
$$
\eqalign{
V_{i+1} = -i \sum_{\alpha_1 + 2\alpha_2 + 4\alpha_4 + 5\alpha_5 + \cdots
= i} (-)^{\alpha_1 + \alpha_2 + \alpha_4 +\alpha_5 +\cdots}
 {(\partial_{t_1})^{\alpha_1} \over \alpha_1!} {({1\over 2}\partial
_ {t_2})^{\alpha_2} \over \alpha_2!} {({1\over 4}\partial_{t_4})
^{\alpha_4}\over \alpha_4!} & {({1\over 5}\partial_{t_5})^{\alpha_5}
\over \alpha_5!}\cr
&\cdots \log\tau\cr}
\eqno(3.22)
$$
For $n = 0$, we get,
$$
res~(M L^{1\over 3}) = 0 = {1\over 3}\sum_{n=1\atop n \not= 0(mod 3)}
n t_n~res~L^{n\over 3} + {1\over 3} \sum_{i=1}^\infty V_{i+1}~res~
L^{-{i\over 3}}
\eqno(3.23)
$$
The second term will contribute only for $i=1$ and therefore, (3.23) can be
expressed after integrating with respect to $x$ and multiplying by
$\tau$ as,
$$
\left[\sum_{k=0\atop i=1,2}^\infty (k+{i\over 3}) t_{3k+i} {\partial \over
\partial t_{3k+i}} + C\right] \tau = 0
\eqno(3.24)
$$
where `$C$' is an integration constant which can not be fixed at this
stage.

For $n\geq 1$, using the expression of $M$ in (3.21) we get from
$res~~(M L^{n+{1\over 3}}) = 0$ after an integration with respect to
$x$ and multiplying by $\tau$,
$$
\left[\sum_{k=0 \atop i=1,2}^\infty (k+{i\over 3}) t_{3k+i} {\partial \over
\partial t_{3(k+n)+i}} + {1\over 6} \sum_{ i+j = 3n \atop i,j \ne 0(mod 3)}
{\partial^2 \over \partial t_i \partial t_j} \right] \tau = 0
\eqno(3.25)
$$
The functions $V_{i+1}$ in this case satisfy the identity
$$
\sum_{k=0 \atop i=1,2} ^n V_{3k+i+1}~ res~ L^{3n-3k-i \over 3} = {\partial
\log\tau \over \partial t_{3n+i}} + {1\over 2} {\partial \over
\partial x} \left[{1 \over \tau} \sum_{ i+j=3n \atop i,j \ne 0(mod 3)}
{\partial^2 \tau
\over \partial t_i \partial t_j} \right]~~~~~~~~~n \ge 0
\eqno(3.26)
$$
The eqs. (3.15), (3.24) and (3.25) can be written combinedly as,
$$
L_n \tau = 0 \qquad\qquad n\ge -1
\eqno(3.27)
$$
If we now impose the condition that $L_n$ would satisfy a centerless
Virasoro algebra, then the integration constant in (3.24) can be fixed
to be ${ 1\over 9}$ and the forms of the Virasoro generators would be
$$
\eqalign{ L_{-1} &= \sum_{k=1 \atop i=1,2}^\infty (k+{i\over 3})
t_{3k+i} {\partial \over \partial t_{3(k-1)+i}} + {2\over 3} xy\cr
L_0 &= \sum_{k=0 \atop i=1,2}^\infty (k+{i\over 3}) t_{3k+i} {\partial
\over \partial t_{3k+i}} + {1\over 9}\cr
L_n &= \sum_{k=0 \atop i=1,2}^\infty (k+{i\over 3}) t_{3k+i} {\partial
\over \partial t_{3(k+n)+i}} + {1\over 6} \sum_{i+j=3n \atop i,j \ne
0(mod 3)} {\partial^2 \over \partial t_i \partial t_j} \qquad n\ge 0\cr}
\eqno(3.28)
$$
In order to show that $\tau$-function of the Boussinesq hierarchy also
satisfies $W^{(3)}$ constraints, we will make use of the relations (3.19)
and (3.20). First, we note that the relation (3.18a) implies
$$
[ M, L^{n+{1\over 3}}] = -(n+{1\over 3}) L^n
\eqno(3.29)
$$
Therefore,
$$
[(M L^{ n+{1\over 3}})~(M L^{m+{1\over 3}})]_- = (M^2 L^{m+n+{2\over 3}})_-
= 0 ~~~~~~for~~~ m\geq 0, n\geq 0
\eqno(3.30)
$$
For $m+n = -1$, we can show using (3.19) that,
$$
res~~ (M L^{-{1\over 3}}) = 0
\eqno(3.31)
$$
and finally we have using (3.19) and (3.18c)
$$
\left(M^2 L^{-{4\over 3}} - (2\alpha +{2\over 3}) M L^{-{5\over 3}} + (\alpha
+ \alpha^2) L^{-2} \right)_- = 0
\eqno(3.32)
$$
We would like to point out here that the particular combination (3.32)
produces correct $W_{-2}^{(3)}$ constraint only for
$\alpha = { 1\over 3}$ [24].
As we mentioned in the KdV hierarchy case, that $\alpha$ can not be fixed
from the relations (3.18) and the string equation (3.16). It can be fixed
to this particular value ${1\over 3}$, by making use of the residual symmetry
of the Zakharov-Shabat dressing operator as noted in ref.[34].

The operator $M^2$ can be expressed as a power series in the Boussinesq Lax
operator (see appendix) as,
$$
\eqalign{ M^2 &= {1\over 9} \left[\sum_{{n=1} \atop
{n\not= 0(mod 3)}}^\infty
n(n-1) t_n L^{n-2 \over 3} + \sum_{{n,m = 1} \atop {n,m \not= 0(mod 3)}}
^\infty n m t_n t_m L^{n+m-2 \over 3}\right.\cr
&\qquad\qquad + 2\sum_{{n,i =1}\atop {n\not= 0(mod 3)}}^\infty
n t_n V_{i+1} L^{n-i-2 \over 3}
-\sum_{i=1}^\infty (i+1) V_{i+1} L^{-i-2 \over 3}\cr
 &\qquad\qquad\qquad\left.  + \sum_{i,j = 1}^\infty
V_{i+1} V_{j+i} L^{-i-j-2 \over 3} \right] \cr}
\eqno(3.33)
$$
Using the expression of $M$ in (3.21) and $M^2$ in (3.33), it is quite
straightforward to calculate the residue of (3.32) and we get after an
integration with respect to $x$ and then multiplying by $\tau$
$$
{1\over 9}\left[ \sum_{m+n-p = 6 \atop m,n,p \ne 0(mod 3)} n m t_n t_m
{\partial \over \partial t_p} + \sum_{m-n-p = 6 \atop m,n,p \ne 0(mod 3)}
m t_m {\partial^2 \over \partial t_n \partial t_p} + 4 t_1^2 t_4 + C_6
\right] \tau = 0
\eqno(3.34)
$$
where $C_6$ is an integration constant which does not depend on $x$ and
has scaling dimension 6. We have also made use of the expression of
$V_{i+1}$ given in (3.22) and noted that they satisfy the following
relation
$$
\eqalign{&
\sum_{n \ge 7 \atop n \ne 0(mod 3)} n t_n \left[(n-5){ \partial^2 \log \tau
\over \partial x \partial t_{n-6}} - 2 {\partial \log \tau \over \partial
t_{n-5}} + 2 V_{n-5}\right] + 2 \sum_{n-m \ge 7 \atop n \ne 0(mod 3)} n t_n
V_{m+1} {\partial^2 \log \tau \over \partial x \partial t_{n-m-6}}\cr
& \qquad\qquad = {\partial \over \partial x} \left[ {1\over \tau}
\sum_{m-n-p = 6\atop
m,n,p \ne 0(mod 3)} m t_m {\partial^2 \over \partial t_n \partial t_p}
\tau\right]\cr}
\eqno(3.35)
$$
Similarly, we calculate for $p = -1$
$$
\eqalign{res~(M^2 L^{p+ {2\over 3}})&= {1\over 9}\left[\sum_{n=1 \atop
n \ne 0(mod 3)}^\infty n(n-1) t_n L^{n-3 \over 3} + \sum_{n,m = 1 \atop
n,m \ne 0(mod 3)}^\infty n m t_n t_m L^{n+m-3 \over 3}\right.\cr
& \qquad\qquad\left. + 2\sum_{n,m = 1 \atop n \ne 0(mod 3)}^\infty
n t_n V_{m+1} L^{n-m-3 \over 3} \right] \cr}
\eqno(3.36)
$$
Again after integration with respect to $x$ and multiplying by $\tau$
the above expression reduces to,
$$
{1\over 9}\left[\sum_{m+n-p = 3 \atop m,n,p \ne 0(mod 3)} n m t_n t_m {
\partial \over \partial t_p} + \sum_{m-n-p = 3 \atop m,n,p \ne 0(mod 3)}
m t_m {\partial^2 \over \partial t_n \partial t_p} + {1\over 3} x^3 +
C_3\right]
\tau = 0
\eqno(3.37)
$$
Here $C_3$ is the integration constant independent of $x$ and has scaling
dimension 3. We have used the explicit form of $V_{i+1}$ and made use of
the first Virasoro constraint $L_{-1} \tau = 0$. Proceeding in a similar way
for higher values of $p$, we recover the higher $W^{(3)}$ constraints in the
form
$$
W_n^{(3)} \tau = 0 \qquad\qquad n\geq 0
$$
where
$$
W_n^{(3)} = {1\over 9} \left[\sum_{p+q-r = 3n
\atop p,q,r \ne 0(mod 3)} p q t_p t_q
{\partial \over \partial t_r} + \sum_{p-q-r = 3n \atop p,q,r \ne 0(mod 3)}
p t_p {\partial^2 \over \partial t_q \partial t_r} + {1\over 3}\sum_{p+q+r
= -3n} {\partial^3 \over \partial t_p \partial t_q \partial t_r}\right]
\eqno(3.38)
$$
In obtaining this we have to use the Virasoro constraints and also we note
the simple identity of the form
$$
\eqalign{{\partial^3 \log \tau \over \partial t_m \partial t_n \partial
t_p} &+ ({\partial^2 \log \tau \over \partial t_m \partial t_n}{\partial
\log \tau \over \partial t_p} + {\partial^2 \log \tau \over \partial t_n
\partial t_p}{\partial \log \tau \over \partial t_m} + {\partial^2 \log \tau
\over \partial t_m \partial t_p}{\partial \log \tau \over \partial t_n})\cr
&\qquad +{\partial \log \tau \over \partial t_m}
{\partial \log \tau \over \partial
t_n}{\partial \log \tau \over \partial t_p} = { 1\over \tau}({\partial^3 \tau
\over \partial t_m \partial t_n \partial t_p})\cr}
\eqno(3.39)
$$
Note that we have conditions $p,q,r \ne 0(mod 3)$ unlike in ref.[34], because
in our definition of the operator $M$, we have removed the coordinates
$t_{3k}$, $k= 1,2, \ldots$. Eqs.(3.34), (3.37) and (3.38) can be written
combinedly in the form
$$
W_n^{(3)} \tau = 0 \qquad\qquad n \geq -2
\eqno(3.40)
$$
It is easy to check that the constraints $L_n \tau = 0$ for $n \ge -1$ and
$W_n^{(3)} \tau = 0$ for $n \ge -2$ satisfy a closed $W^{(3)}$-algebra
provided the integration constant appearing in (3.36) is ${8\over 27} t^3_2$
and $C_3$ in (3.37) is zero. In that case, we have the standard matrix-model
form of the $W^{(3)}$ constraints
$$
\eqalign{W_n^{(3)} &= {1\over 9}\left[{1\over 3}
\sum_{p+q+r = 3n} p q r t_p t_q t_r +
\sum_{p+q-r = 3n} p q t_p t_q {\partial \over \partial t_r}
+ \sum_{p-q-r= 3n} p t_p {\partial^2 \over \partial t_q \partial t_r}
\right. \cr
& \qquad\qquad\left. + {1\over 3}\sum_
{p+q+r = -3n}{\partial^3 \over \partial t_p \partial t_q \partial t_r}
 \right] \qquad\qquad
n \geq -2 \cr}
\eqno(3.41)
$$
and $p,q,r$ in all the terms are not $0(mod 3)$.
\bigskip
{\titlea IV. Generalization to Higher KdV Hierarchies:}
\bigskip
Generalization of the method described in sections II and III can be most
easily done if we note that the operator $M$ for the $p$-reduced KP hierarchy
satisfies
$$
[L^{1\over p}, M] = {1\over p}
\eqno(4.1)
$$
The Lax operator for the $p$-reduced KP hierarchy is defined as
$$
L = \partial^p + u_{p-2}\partial^{p-2} + \cdots + u_0
\eqno(4.2)
$$
and the additional symmetry operator $M$ has the form
$$
M = {1\over p} K ( \sum_{n=1 \atop n \ne 0(mod p)}^\infty n t_n \partial
^{n-1}) K^{-1}
\eqno(4.3)
$$
Douglas' string equation for the general massive model in this case can be
written as,
$$
\sum_{k=1 \atop i=1,2,\ldots, p-1}^\infty (k+{i\over p}) t_{kp+i}~
[ L, L^{p(k-1)+i \over p}_+] = 1
\eqno(4.4)
$$
Since the operators $L$ and $M$ satisfies the fundamental relation (4.1),
we can show that
$$
[ L, ML^{{1\over p}-1}] = 1
\eqno(4.5)
$$
Using the definition of $M$ in (4.3), it is an easy exercise to check that
(4.4) can also be written equivalently as,
$$
[ L, (ML^{{1\over p}-1})_+] = 1
\eqno(4.6)
$$
The string equation (4.6) and the relation (4.5) together therefore implies,
$$
[ L, (ML^{{1\over p}-1})_-] = 0
\eqno(4.7)
$$
It is therefore clear that the operator $(ML^{{1\over p}-1})_-$ should be
some negative powers of $L$. Using the relation (4.1) we can also show that
$$
[ M, L^{{1\over p}-1} ] = {p-1 \over p} L^{-1}
\eqno(4.8)
$$
{}From (4.8) we conclude that
$$
(ML^{{1\over p}-1})_- = \alpha L^{-1}
\eqno(4.9)
$$
where $\alpha $ is an arbitrary constant. Evaluating the residue of (4.9)
we can easily show that this is equivalent to the string equation (4.4)
or (4.6). Eq.(4.9) gives an infinite set of relations of the form
$$
(ML^{n+{1\over p}})_- = ((ML^{{1\over p}-1})_- L^{n+1})_- = 0
{}~~~~~~n\ge 0
\eqno(4.10)
$$
The cosequences of (4.9) and (4.10) together can be worked out using the
expression of $M$ in (4.3) and the Lax operator (4.2) of the $p$-reduced
KP hierarchy. Calculating the residue and after an integration with respect
to $x$ and then multiplying by $\tau$, they can be shown to be equivalent
to the semi-infinite set of Virasoro constraints following the discussions
in sections II and III,
$$
L_n \tau = 0~~~~~~~~~~~~n\ge -1
\eqno(4.11)
$$
where $L_n$'s would have the form
$$
\eqalign{L_n = {1\over 2p}\sum_{i+j = -pn}& i j t_i t_j
+ {1\over p}\sum_{i-j = -pn}
i t_i {\partial \over \partial t_j}\cr
&\qquad + {1\over 2p}\sum_{i+j = pn}
{\partial^2 \over \partial t_i \partial t_j} + {p^2 - 1 \over 24p}
\delta_{n,-1}\cr}
\eqno(4.12)
$$
for $n\geq -1$ and $i,j$ do not take values $0(mod p)$.

In order to get higher $W$-constraints, one has to take various powers of
the operator $(ML^{{1\over p}-1})$. They could be simplified using the basic
relation (4.1) as explained for the case of $W^{(3)}$ in section III. Also,
we point out that $\alpha$ in (4.9) could remain arbitrary in order to get
correct Virasoro constraints as we just explained but this no longer remains
to be true for obtaining $W$ constraints. We saw in section III that the
constant has to be fixed to a particular value for obtaining correct $W^{(3)}$
constraints. In this case the constant has to be chosen as ${p-1\over 2p}$.
This is usually done by using the residual symmetry of the opeartor $K$ noted
in ref.[34]. In order to obtain higher powers of $M$ as a power series
expansion of the Lax oparator, we make use of the relation
$$
M \psi = {1\over p} {\partial \psi \over \partial \lambda}
\eqno(4.13)
$$
where $\psi$ is known as the Baker-Akhiezer function of the $p$-reduced KP
hierarchy [26] defined as
$$
\psi = K~~ exp~ (\sum_{n=1 \atop n\ne 0(mod p)} t_n \lambda^n)
\eqno(4.14)
$$
and $\lambda$ is the spectral parameter. We note that the Baker-Akhiezer
function for $p$-reduced KP hierarchy does not depend on the coordinates
$t_{kp}$, $k=1,2,\ldots$. By successively applying $M$ to (4.13) one can
obtain
$$
M^n \psi = {1\over p^n}{\partial^n \psi \over \partial \lambda^n}~~~~~~~~
n<p
\eqno(4.15)
$$
One can easily express $({1\over \psi} {\partial^n \psi \over \partial
\lambda^n})$ in terms of $\log \psi$ using the recursion relation of the form
$$
A_n = {\partial A_{n-1} \over \partial \lambda} + {\partial \log \psi \over
\partial \lambda} A_{n-1}~~~~~~n\ge 1
\eqno(4.16)
$$
where we defined $A_n \equiv {1\over \psi}{\partial^n \psi \over \partial
\lambda^n}$. Expressing $\log \psi$ in powers of $\lambda$ as,
$$
p\log \psi = \sum_{n=1\atop n\ne 0(mod p)}^\infty t_n \lambda^n
-\sum_{i=1}^\infty {1\over i} V_{i+1} \lambda^{-i}
\eqno(4.17)
$$
and then plugging (4.17) and (4.16) in (4.15) we obtain an
expression of $M^n$ in power
series of $L$ after replacing $\lambda$ by $L^{1\over p}$. With this procedure,
it is quite strightforward to calculate various powers of
$(ML^{{1\over p}-1})$. Thus using the procedure described in sections
II and III, we can get the full set of $W^{(p)}$ constraints.
\vskip .5cm
{\titlea V. Conclusions:}
\vskip .2cm
By expanding the additional symmetry operator associated with the $p$-reduced
KP hierarchy as a power series in the Lax operator we have shown directly
that the Virasoro and the $W$-constraints do indeed follow from the string
equation for such integrable systems. Our method also clarifies the reason
for the appearance of the constant term in $L_0$ constraint and the appearance
of $t_2^3$ term in $W_{-2}^{(3)}$ constraint which are not clear in other
approaches. We noted that these terms appear as integration constants and can
be fixed from the algebra in analogy with the matrix model results. We have
emphasized that the correct Virasoro constraints may be obtained without any
restriction on the additional symmetry operator $M$, but the correct $W$-
constraints follow only if we put an additional condition ${\partial M
\over \partial t_{kp}} = 0,~~~k=1,2,\ldots$. We have noticed that the
generators
of the additional symmetries $(M^n L^{m+{n\over p}}),~~~~~1\le n\le p-1,
{}~~~~~m\ge -1$ give rise to the Virasoro and the $W$-constraints if we
consider
only the residue of this operator. There are infinite other conditions
corresponding to the coefficients of $\partial^{-2}$, $\partial^{-3}, \ldots$
terms which remain unexplored. It would be interesting to see whether they
correspond to further constraints. It is possible that they are nothing but
the Virasoro and $W$-constraints in various guises. Also. it is not clear
what sort of constraints one would get if one considers the operators
$(ML^{m+{n\over p}})$ for $n\ge p$. These questions are presently under
investigation.
\vskip 1cm
{\titlea Acknowledgements:}
\bigskip
The authors would like to thank Professor Abdus Salam, International
Atomic Energy Agency and UNESCO  at the International Centre for
Theoretical Physics, Trieste for hospitality and support.
\vfill
\eject
\centerline{\titlea Appendix}
\bigskip
Here we list a few useful formulas for the KP hierarchy. The corresponding
formulas for $p$-reduced KP hierarchy which were used in the text can be
obtained easily using the reduction condition $(L_{KP}^p)_- = 0$ and
${\partial \tau \over \partial t_{kp}} = 0$ for $k=1,2,\ldots$. Here
$L_{KP}$ is the KP Lax operator given as
$$
L_{KP} = \partial + \sum_{i=1}^\infty u_{i+1}(t) \partial^{-i}
\eqno(A.1)
$$
where $t$ denotes the infinite set of evolution parameters $(t_1, t_2,
t_3,\ldots)$ and $\tau$ is the $\tau$-function of the $p$-reduced KP
hierarchy. The KP hierarchy is described in terms of the Lax equation
$$
{\partial L_{KP} \over \partial t_n} = [ (L_{KP}^n)_+, L_{KP}]
\eqno(A.2)
$$
{}From the Lax equation $t_1$ can be identified with $x$. The Baker-Akhiezer
function $\psi$ satisfies the eigenvalue equation
$$
L_{KP} \psi = \lambda \psi
\eqno(A.3)
$$
and has a solution of the form
$$
\psi(t) = K(t)~exp~(\sum_{n=1}^\infty t_n \lambda^n)
\eqno(A.4)
$$
Where the Zakharov-Shabat dressing operator is defined as
$$
K(t) = 1 + \sum_{i=1}^\infty a_i \partial^{-i}
\eqno(A.5)
$$
The additional symmetry operator $M$ is defined as
$$
\eqalign{M &= K ( \sum_{n=1}^\infty n t_n \partial^{n-1}) K^{-1}\cr
&= \sum_{n=1}^\infty n t_n L^{n-1} + \sum_{i=1}^\infty V_{i+1} L^{-i-1}\cr}
\eqno(A.6)
$$
where $V_{i+1} = -(i a_i + \sum_{j=1}^{i-1} j a_j {\tilde a}_{i-j})$. The first
two $V$'s have the form
$$V_2 = -a_1
\eqno(A.7a)
$$
$$
V_3 = -2 a_2 + a_1^2
\eqno(A.7b)
$$
The operators $L_{KP}$ and $M$ satisfies
$$
[L_{KP}, M] = 1
\eqno(A.8)
$$
{}From the form of $M$ in (A.6) we derive,
$$
M \psi = {\partial \psi \over \partial \lambda}
\eqno(A.9)
$$
which gives
$$
\sum_{n=1}^\infty n t_n \lambda^{n-1} + \sum_{i=1}^\infty V_{i+1} \lambda^
{-i-1} = {\partial \log \psi\over \partial \lambda}
\eqno(A.10)
$$
In terms of the $\tau$-function
$$
\psi = {\left(exp(\sum_{n=1}^\infty t_n \lambda^n) - exp(-\sum_{n=1}^\infty
{1\over n} \lambda^{-n} {\partial \over \partial t_n})\right) \tau(t)
\over \tau(t)}
\eqno(A.11)
$$
So,
$$
\log \psi = \sum_{n=1}^\infty t_n \lambda^n + \sum_{N=1}^\infty {1\over N!}
\left(-\sum_{n=1}^\infty {1\over n}
\lambda^{-n} {\partial \over \partial t_n} \right)^N
\log \tau(t)
\eqno(A.12)
$$
Comparing (A.12) and (A.10) we obtain
$$
V_{i+1} = -i \sum_{\alpha_1+2\alpha_2+3\alpha_3+\cdots = i} (-)^{\alpha_1
+\alpha_2+\alpha_3+\cdots} {(\partial_{t_1})^{\alpha_1}\over \alpha_1!}
{({1\over 2}\partial_{t_2})^{\alpha_2} \over \alpha_2!} {({1\over 3}\partial
_{t_3})^{\alpha_3}\over \alpha_3!}\cdots \log\tau(t)
\eqno(A.13)
$$
The first two $V$'s calculated from above have the form
$$V_2 = {\partial \log \tau \over \partial t_1}
\eqno(A.14a)
$$
$$
V_3 = {\partial \log \tau \over \partial t_2} - {\partial^2 \log \tau
\over \partial t_1^2}
\eqno(A.14b)
$$
{}From the evolution equation (A.2), we get the evolution equation for the
Zakharov-Shabat dressing operator as
$$
{\partial K\over \partial t_n} = - (L_{KP}^n)_- K
\eqno(A.15)
$$
where we made use of the fact that the Zakharov-Shabat dressing operator
satisfies
$$
L_{KP} = K \partial K^{-1}
\eqno(A.16)
$$
{}From (A.15) we get
$$
{\partial a_1 \over \partial t_n} = - res~~L_{KP}^n
\eqno(A.17a)
$$
and
$$
{\partial a_2 \over \partial t_n} = - (res~~L_{KP}^n) a_1 - (L_{KP}^n)_{-2}
\eqno(A.17b)
$$
Since $V_2 = a_1 = {\partial \log\tau \over \partial t_1}$, so, from (A.17a)
we get
$$
{\partial^2 \log\tau \over \partial t_1 \partial t_n} = res~~L_{KP}^n
\eqno(A.18)
$$
Similarly from (A.7b)
$$
a_2 = -{1\over 2}{\partial\log\tau \over \partial t_2} + {1\over 2}{\partial
^2 \log\tau \over \partial t_1^2} + {1\over 2}({\partial \log \tau \over
\partial t_1})^2
\eqno(A.19)
$$
Using (A.19) in (A.17b) we get,
$$
{\partial^2 \log\tau \over \partial t_2 \partial t_n} = (res~~L_{KP}^n)'
+2 (L_{KP}^n)_{-2}
\eqno(A.20)
$$
We also show how $M^2$ can be calculated using (A.9). Since $M^2\psi =
{\partial^2\psi \over \partial \lambda^2}$ so,
$$
M^2 \psi = \left[ {\partial^2 \log\psi \over \partial \lambda^2} + ({\partial
\log\psi \over \partial \lambda})^2\right]\psi
\eqno(A.21)
$$
Using (A.10) we get,
$$
\eqalign{ M^2\psi &= \left[\sum_{n=1}^\infty n(n-1)
t_n \lambda^{n-2} + \sum_
{i=1}^\infty (-i-1) V_{i+1} \lambda^{-i-1}\right. \cr
&\left. \quad + (\sum_{n=1}^
\infty n t_n \lambda^{n-1} + \sum_{i=1}^\infty V_{i+1}
\lambda^{-i-1})~(\sum_{m=1}^\infty m t_m \lambda^{m-1} + \sum_{j=1}^\infty
V_{j+1} \lambda^{-j-1}) \right] \cr}
\eqno(A.22)
$$
So, in terms of the Lax operator $M^2$ can be expressed as,
$$
\eqalign{ M^2 &= \sum_{n=1}^\infty n(n-1) t_n L_{KP}^{n-2} + \sum_{i=1}
^\infty (-i-1) V_{i+1} L_{KP}^{-i-2} +
\sum_{n,m =1}^\infty n m t_n t_m L_{KP}^{n+m-2}\cr
&\qquad\qquad + 2\sum_{n,i =1}^\infty n t_n V_{i+1}
L_{KP}^{n-i-2} + \sum_{i,j=1}^\infty
V_{i+1} V_{j+1} L_{KP}^{-i-j-2}\cr}
\eqno(A.23)
$$
\vfill
\eject
\vskip .5cm
{\titlea References:}
\bigskip
\item{1.} D. Gross and A. Migdal, Phys. Rev. Lett. 64 (1990), 127; Nucl. Phys.
B340 (1990), 333.
\item{2.} M. Douglas and S. Shenkar, Nucl. Phys. B335 (1990), 635.
\item{3.} E. Brezin and V. Kazakov, Phys. Lett. B236 (1990), 144.
\item{4.} E. J. Martinec, Univ. of Chicago preprint EFI-90-67, (1990).
\item{5.} L. Alvarez-Gaume, CERN preprint CERN-TH-6123/91 (1991).
\item{6.} R. Dijkgraaf and E. Witten, Nucl. Phys. B342 (1990), 486.
\item{7.} E. Verlinde and H. Verlinde, Nucl. Phys. B348 (1991), 457.
\item{8.} J. Distler, Nucl. Phys. B342 (1990), 523.
\item{9.} E. Witten, Surveys In Diff. Geom. 1 (1991), 243; preprint
IASSNS-HEP-91/74.
\item{10.} R. Myers and V. Periwal, Nucl. Phys. B333 (1990), 536.
\item{11.} M. Kontsevich, Max-Planck Institute preprint MPI/91-47; MPI/91-77.
\item{12.} S. Chadha, G.Mahoux and M. L. Mehta, J. Phys. A14 (1981), 579.
\item{13.} M. R. Douglas, Phys. Lett. B238 (1990), 176.
\item{14.} P. D. Lax, Comm. Pure Appl. Math. 21 (1968), 467; ibid 28 (1975),
141.
\item{15.} L. D. Fadeev and L. A. Takhtajan, \lq\lq Hamiltonian methods in the
Theory of Solitons", Springer-Verlag (1987).
\item{16.} A. Das, \lq\lq Integrable Models", World Scientific (1989).
\item{17.} H. S. La, Comm. Math. Phys. 140 (1991), 569.
\item{18.} G. Moore, Comm. Math. Phys. 133 (1990), 261.
\item{19.} V. Kac and A. Schwartz, Phys. Lett. B257 (1991), 329.
\item{20.} M. Fukuma, H. Kawai and R. Nakayama, Comm. Math. Phys. 143 (1992),
371.
\item{21.} K. N. Anagnostopoulos, M. J. Bowick and A. Schwartz, Syracuse
preprint SU-4238-497 (1991).
\item{22.} M. Fukuma, H. kawai and R. Nakayama, Int. Jour. Mod. Phys. A6
(1991), 1385.
\item{23.} E. Gava and K. S. Narain, Phys. Lett. B263 (1991), 213.
\item{24.} J. Goeree, Nucl. Phys. B358 (1991), 737.
\item{25.} E. Date, M. Kashiwara, M. Jimbo and T. Miwa in ``Nonlinear
Integrable Systems" eds. M. Jimbo and T. Miwa, World Scientific (1983).
\item{26.} G. Segal and G. Wilson, Publ. Math. IHES 63 (1985), 1.
\item{27.} S. Panda and S. Roy, ICTP preprint IC/92/145.
\item{28.} N. K. Ibragimov and A. B. Shabat, Sov. J. Phys. Dokl. 244 (1979),
57.
\item{29.} A. Y. Orlov and E. I. Shulman, Lett. Math. Phys. 12 (1986), 171.
\item{30.} P. G. Grinevich and A. Y. Orlov, in ``Problems of Modern Quantum
Field Theories", eds. A. A. Belavin, A. U. Klymik and A. B. Zamolodchikov
(Springer-Verlag), 1989.
\item{31.} L. A. Dickey, Univ. of Oklahoma preprint (1992).
\item{32.} K. Takasaki and T. Takebe, preprint RIMS-814, 1991.
\item{33.} A. Das and S. Roy, Int. Jour. Mod. Phys. A6 (1991), 1429.
\item{34.} M. Adler and P. Van-Moerbeke, Brandeis preprint (1992).
\end